**Title:** There are no equal opportunity infectors: Epidemiological modelers must rethink our approach to inequality in infection risk.

**Short Title:** Rethinking the role of social inequality in transmission models.

**Authors:** Jon Zelner[1,2], Nina B. Masters[1], Ramya Naraharisetti[1,2], Sanyu Mojola[3], Merlin Chowkwanyun[4], Ryan Malosh[1]

1. Dept. of Epidemiology, University of Michigan School of Public Health
2. Center for Social Epidemiology and Population Health, University of Michigan School of Public Health
3. Dept. of Sociology, School of Public and International Affairs & Office of Population Research, Princeton University
4. Dept. of Sociomedical Sciences, Mailman School of Public Health, Columbia University

**Abstract**. Mathematical models have come to play a key role in global pandemic preparedness and outbreak response: helping to plan for disease burden, hospital capacity, and inform non-pharmaceutical interventions. Such models have played a pivotal role in the COVID-19 pandemic, with transmission models – and, by consequence, modelers – guiding global, national, and local responses to SARS-CoV-2. However, these models have systematically failed to account for the social and structural factors which lead to socioeconomic, racial, and geographic health disparities. Why do epidemiologic models of emerging infections ignore known structural drivers of disparate health outcomes? What have been the consequences of this limited framework? What should be done to develop a more holistic approach to modeling-based decision-making during pandemics? In this Perspective, we evaluate potential historical and political explanations for the exclusion of drivers of disparity in infectious disease models for emerging infections, which have often been characterized as "equal opportunity infectors" despite ample evidence to the contrary. We look to examples from other disease systems (HIV, STIs) as a potential blueprint for how social connections, environmental, and structural factors can be integrated into a coherent, rigorous, and interpretable modeling framework. We conclude by outlining principles to guide modeling of emerging infections in ways that represent the causes of inequity in infection as central rather than peripheral mechanisms.

**Introduction**

In March 2020, a prescient news item in *Science* proclaimed that infectious disease transmission models had taken on "life or death importance"[1] as tools in the fight against SARS-CoV-2. Despite the pivotal role transmission models and modelers have played, most models used to guide the global response to SARS-CoV-2 paid virtually no attention to the causes of the massive socioeconomic and racial inequities that have characterized the pandemic[2–4]. As ongoing domestic and global vaccination campaigns and the emergence of novel and more infectious variants continue to highlight new facets of inequity in the COVID-19 crisis, recent comprehensive reviews of the transmission modeling literature have begun to grapple with how transmission models can and should account for 'social factors' going forward[5,6]. These reviews highlight the increased use of detailed information on mobility and individual behavior to make better predictions of epidemic trajectories and estimates of model parameters. In broad strokes, they argue for more detailed data collection and closer partnership with communities to ensure that models incorporate local data and address real-world needs.

These are critically important points. However, more and better data – even when collected and used in partnership with affected communities – does not guarantee models that address the roots of infection inequity. In this piece, we advocate for an approach to infectious disease modeling that explicitly includes overarching social factors, like socioeconomic status (SES), racism, and segregation that put individuals and populations "at risk of risk"[7]. We do not in any way eschew the use of detailed, high-resolution social data as a tool for combating infectious disease. Instead, we worry that the power of these tools to prevent and ameliorate inequity may be squandered if their use is not informed by key concepts from the vast literature on health disparities in both infectious and non-communicable disease: When structural causes are excluded from transmission models, they cannot be used to examine how structural remedies, like wealth transfers, universal healthcare, and guaranteed housing might impact incidence and mortality rates.

These omissions are not the fault of individual authors, and no model can be expected to capture the full gamut of social and biological causes of infection. Instead, the absence of models that account for these higher-order factors reflects a lack of a digestible framework for including social inequality as a first-class feature in transmission models. The current moment provides an opportunity to rectify this gap. COVID-19 has increased awareness of the fact that flesh-and-blood social inequities underlie the values of abstract model parameters. For example, Richardson et al. have argued that, in the United States, the basic reproduction number, $R_0$, must be understood not only in terms of pathogen biology and individual behavior, but also of structural violence. Various axes of difference – race and class among them -- compel some to be routinely and non-randomly exposed and allows others to remain relatively safe.[8] Similarly, the social epidemiologist David Williams has argued that herd immunity should be reconceptualized in explicitly social terms which recognize that the level of immunity from natural infection and vaccination is a function not only of pathogen biology, but the social and economic systems that propel transmission and health behavior.[9]

While structural inequalities have been largely absent from models of ARIs such as SARS-CoV-2 and pandemic influenza, they have long been at the heart of modeling work on HIV as well as a range of viral and bacterial sexually transmitted infections (STIs). Because transmission of STIs typically requires direct contact, interpersonal relationships and the social factors that structure them cannot be ignored, and there is a vast literature examining inequities in HIV and STI infection using population- and network-based transmission models[10–12]. By contrast, transmission of ARIs through the air via respiratory droplets and aerosols makes mapping of social relationships onto transmission less straightforward, although smaller-scale studies have demonstrated the role of individual-level patterns of contact in influenza transmission[13]. This disconnect has contributed to the erroneous, but pervasive, idea that ARIs are "equal opportunity infectors"[14] for which the structure of social networks and systems of both global and domestic inequality and oppression are less important than for quintessentially social pathogens like HIV, STIs, and TB. COVID-19 has dealt a significant blow to this idea, but it remains to be seen how we can capitalize on that momentum to make the necessary changes to the modeling toolkit.

The way we represent cause and effect in transmission models has enormous implications for policy and practice. Transmission models let practitioners, policymakers, and researchers envision alternative futures that might be realized through intervention, policy, and social action. When these models exclude social-structural factors that drive inequity - income, education, racial residential segregation, and occupational hazards - they preclude the ability to explore the possibility of structural change as an epidemiological tool on par with NPIs, vaccination, and testing. In the following sections, we outline an "equity-forward" approach to transmission modeling that places the fundamental structural causes of infection inequality on an equal level with the biological and behavioral features of transmission. Our goal is to articulate a vision of socially informed modeling that is squarely focused on understanding how imbalances in social power drive infection inequalities and suggesting social and political remedies to these disparities.

**What are the goals of equity-forward transmission modeling?**

Transmission models are often used to answer 'what-if' questions from a perspective of power or authority: What will happen if governments impose quarantines or mask-wearing orders? How should national, state and local public health authorities allocate scarce vaccines? These questions lend themselves to a focus on predicting the timing and spatial distribution of infection over short and long time horizons[5]. This focus on predicting the future is most relevant for a decision-maker who is anticipating, planning, and responding to events in the near future.[15] This may explain the sustained focus on developing epidemic forecasting models despite their spotty track record at predicting the future[16]. Recent reviews have argued that the failures of the predictive/forecasting approach may be traceable to missing social data or the omission of behavioral feedbacks, e.g. individuals restricting their contacts in response to rising caseloads, and that this gap can be closed with more data and more-detailed models. But what if this failure instead reflects the unpredictability of the social, political, and economic human substrate of transmission? The COVID-19 crisis has illustrated how the trajectory of world-historic events like pandemics are shaped by a combination of pathogen biology, human social structures, and contingent historical events that confound efforts to anticipate the future with any degree of certainty.

While forecasting may have a role to play in managing an ongoing crisis, we argue that the focus on optimizing models for short-term prediction is at best ancillary to and at worst an active impediment towards pursuing the goal of equity. For infectious disease models to be more useful as tools for addressing inequity, they need to be primarily *diagnostic* and *forensic* in the sense of being tools we use to identify and measure the causes of disparity. Because the remedies to structural inequities are not discrete, one-off interventions, but instead messy and protracted contests over power, equity-forward models must provide evidence and ideas that can propel and support efforts at social and political change. They should draw on historical and contemporary data to provide a map of a potential path from the unjust here to a more-just there. This requires evaluating the impact of social policies and movements on infection outcomes over longer periods of time, and focusing on 'what-if' questions that center on the conditions *prior* to a crisis: What if the burden of racial residential and occupational segregation was eased or eliminated before the next pandemic? What if healthcare was freely available to all during the first days and weeks transmission of a novel respiratory pathogen? How would this change the constellation of downstream factors we typically address one at a time? How would outcomes improve based on this comprehensive, structural change as compared to the more reactive approach that the prospective, forecasting paradigm facilitates?

Rather than a primary focus on predicting the future (i.e. the pace and timing of infection) at a population level, an equity-forward modeling approach should be concerned with characterizing *who* is likely to be infected and how the distribution of infection reflects allocations of economic and social power at the population level. Accomplishing this requires transmission models that simultaneously accommodate social and biological mechanisms of interpersonal dependence at levels that are higher than individual-to-individual interactions. For example, aggressive COVID-19 lockdowns were enabled by the labor of healthcare, retail, delivery, and warehouse workers who continued to provide goods and services to people stuck at home. As a result, these workers bore much of the brunt of early exposure, infection, and death. This means that lower rates of exposure experienced by wealthier individuals and Whites[17] are a direct result of economic and racial power imbalances. For epidemiological models to be faithful representations of the way infection risk occurs, this type of social dependence needs to be understood to be of equal importance to faithfully representing disease system as the specification of the biological states an individual progresses through following infection.

**Defining the key mechanisms and outcomes in an *equity-forward* transmission model.**

We advocate for an adaptation of the *fundamental-cause (FC)* perspective on health inequity to the problem of transmission. Link & Phelan define a fundamental social cause of health inequity as a factor, like socioeconomic status (SES) or racism, that puts individuals and populations "at risk of risks"[7]. The FC approach focuses on how differentials in social power impact access to the material and social resources – money, occupation, housing, medical care, education, prestige – that structure risks of infection and death. There is nothing new in the idea that social factors are causes of infection on par with biological ones. In the early 19th-century, the French social hygienist Rene Villerme documented strong associations among income, working conditions, and health.[18] In the 1950s, the social medicine pioneers René and Jean-Baptiste Dubos referred to the *Mycobacterium Tuberculosis* as a necessary, but insufficient precondition for TB infection, with social and occupational factors ultimately shaping exposure and susceptibility[19,20]. Similarly, public health historian Samuel Kelton Roberts detailed how racial residential segregation drove TB infection and mortality among African-Americans in 20th century Baltimore via impacts on housing, workplace conditions, medical treatment, and public health policies[21]. These mechanisms have been repeatedly articulated in narrative histories, risk-factor analyses, and mixed-methods studies of many infections including Cholera[22] HIV[23,24], and malaria[25,26]. Clouston et al. found that while high-income U.S. counties were the first to see an introduction of SARS-CoV-2 infection, the pace of infection and mortality in these counties quickly slowed through non-pharmaceutical interventions (NPIs) (e.g. work from home, school closures). Meanwhile, rates of infection and death exploded in counties with lower-income and higher proportions of non-White residents where NPIs were less feasible.[1] Given the ubiquity of the FC perspective in explicit and implicit understandings of infectious disease risk, its absence from transmission modeling is surprising. We argue that the FC approach provides a useful set of principles which can guide equity-forward modeling. To conclude this essay, we outline three core tenets of FC theory and highlight how they relate to the mechanisms of infectious disease transmission (see Figure 1), with an eye towards how they could be integrated into transmission models:

**1. Social factors such as SES and racism are fundamental causes of infection because they operate on *multiple* intervening mechanisms that drive transmission, including housing, occupation, healthcare, and others.** The proximal factors that drive exposure and mortality risk are socially correlated: Individuals in high-risk occupations are more likely to live in crowded conditions and have poor access to care and prevention. Due to such residential segregation and occupational clustering in certain higher-risk sectors (e.g. low-wage service work), individuals sharing these risks are also more likely to have high rates of contact with each other, concentrating the impact of these differential risks within marginalized groups[27]. *Transmission models must not ignore the fact that these proximal drivers* *flow from upstream causes, lest they may make overly optimistic projections of the impact of tweaks to individual proximal risk factors and under-estimate the impact of higher-level social interventions that would improve multiple downstream factors simultaneously.*

**2. Protective non-pharmaceutical interventions, policies, and medical innovations reach more-advantaged individuals first, and this access is the cause of deprivation among lower-SES individuals and communities.** The COVID-19 pandemic has exposed how non-pharmaceutical interventions such as social distancing are structured by economic and racial privilege. In the context of an emerging infection, these effects may be even more acute than with many non-communicable diseases: While more-advantaged individuals wait out infection at home, the response evolves, clinical management of infection improves, and case-fatality rates fall, allowing these groups to sidestep the worst effects almost entirely while minoritized and poorer groups take the brunt of infection and death[4,28]. *Including mechanisms of social and economic dependence in exposure, i.e. the ways in which one group's increased exposure facilitates the decreased exposure of another, is essential for transmission models to be useful tools for identifying and mitigating inequity.*

**3. The same structural factors drive inequity across *multiple infectious disease outcomes*.** The social factors that drive risks for one infection are likely to influence others. The immediate toll of COVID-19 mortality has disproportionately affected low-SES and minority communities. However, emerging evidence suggests the risk of a 'double jeopardy' effect: those communities where hospital systems were overwhelmed, already-insufficient primary care fell behind, and where children were unable to keep up with routine immunizations, are now at risk of outbreaks of other vaccine-preventable diseases, such as measles, pertussis and others. In addition, risk factors for COVID-19 strongly overlap with those for infections such as tuberculosis, influenza, fungal infections such Coccidiomycosis,[31] HIV/STIs, and others. Vaccine hesitancy, which has been more pronounced among lower SES populations, and poor access to prevention and care also impact risk for multiple pathogens, and a higher prevalence of co-infections dramatically increases risks for poor outcomes including hospitalization and death. *Consequently, models that account for potential 'knock-on' effects of one set of social causes on multiple disease outcomes – how the risks for SARS-CoV-2 are related to risks for influenza, other ARIs, HIV, and other infections – are critically necessary to assess the full scope of damage associated with upstream inequities.*

**Conclusion**

A deeper integration of social-epidemiological and biological ideas in transmission models is urgently necessary. The challenge of this undertaking should not be under-stated, nor should the substantial contributions of transmission modelers during the COVID-19 pandemic. We echo the argument made by Bertozzi et al. in the early days of the pandemic[29]: rather than stumbling over attempts at hyper-realism, transmission models should focus on characterizing broad trends in inequity, the mechanisms that generate them, and multi-level interventions that might work to ameliorate infection inequities. We also should not pursue a single 'correct' model that includes all these mechanisms and outcomes at once. Instead, equity-forward models should be another element of the epidemiological toolkit, alongside their forecasting and predictive counterparts.

The value of these models comes from their potential to force policymakers to justify their choices in terms of who is protected and who isn't. Models that include the mechanisms of inequity as first-class citizens can be tools for achieving the social herd immunity described by Williams & Cooper[9], but getting there will require a sober reckoning of how far we currently are from it. Without models that simultaneously speak the languages of transmission, evolution, and social stratification, structural changes that come from contesting power will once again be left out of the universe of possibilities that can be explored using transmission models. Including these mechanisms positions models – and modelers – to address questions of health justice early and often, instead of being on the margins. And by taking an active role in centralizing the structural drivers of inequity in risk – and acting upon that knowledge – modelers can lead the way in ensuring a better outcome for all in the next crisis.

**Figures & Tables**

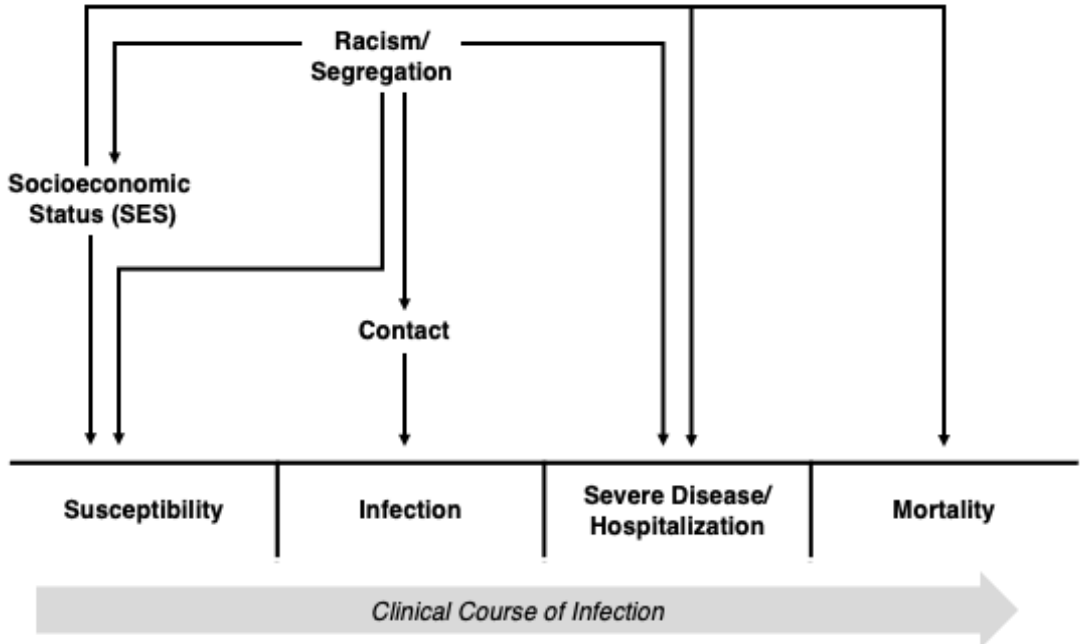

*Figure 1. A conceptual map of the relationship between social-structural inequities and then stages of infection for acute infections such as SARS-CoV-2.* The figure illustrates interdependencies between both sources of social inequality (e.g. racism and socioeconomic status) and their impact on both the rate of infection acquisition and progression of disease through escalating phases of severity.

**References**

1. Enserink M, Kupferschmidt K. With COVID-19, modeling takes on life and death importance. *Science*. 2020;367(6485):1414-1415. doi:10.1126/science.367.6485.1414-b

2. Clouston SAP, Natale G, Link BG. Socioeconomic inequalities in the spread of coronavirus-19 in the United States: A examination of the emergence of social inequalities. *Social Science & Medicine*. 2021;268:113554. doi:10.1016/j.socscimed.2020.113554

3. Zelner J, Trangucci R, Naraharisetti R, et al. Racial disparities in COVID-19 mortality are driven by unequal infection risks. *Clin Infect Dis*. 2020;72(5). doi:10.1093/cid/ciaa1723

4. Kamis C, Stolte A, West JS, et al. Overcrowding and COVID-19 mortality across U.S. counties: Are disparities growing over time? *SSM - Population Health*. 2021;15:100845. doi:10.1016/j.ssmph.2021.100845

5. Bedson J, Skrip LA, Pedi D, et al. A review and agenda for integrated disease models including social and behavioural factors. *Nat Hum Behav*. Published online June 28, 2021:1-13. doi:10.1038/s41562-021-01136-2

6. Buckee C, Noor A, Sattenspiel L. Thinking clearly about social aspects of infectious disease transmission. *Nature*. 2021;595(7866):205-213. doi:10.1038/s41586-021-03694-x

7. Link BG, Phelan J. Social Conditions As Fundamental Causes of Disease. *Journal of Health and Social Behavior*. 1995;35:80. doi:10.2307/2626958

8. Richardson ET, Malik MM, Darity WA, et al. Reparations for Black American descendants of persons enslaved in the U.S. and their potential impact on SARS-CoV-2 transmission. *Social Science & Medicine*. 2021;276:113741. doi:10.1016/j.socscimed.2021.113741

9. Williams DR, Cooper LA. COVID-19 and Health Equity—A New Kind of "Herd Immunity." *JAMA*. 2020;323(24):2478-2480. doi:10.1001/jama.2020.8051

10. Jacquez JA, Simon CP, Koopman JS, Sattenspiel L, Perry T. Modeling and analyzing HIV transmission: the effect of contact patterns. *Mathematical Biosciences*. 1988;92(2):119-199. doi:10.1016/0025-5564(88)90031-4

11. Morris M, Kretzschmar M. Concurrent partnerships and the spread of HIV. *AIDS*. 1997;11(5):641-648.

12. Adams JW, Lurie MN, King MRF, et al. Potential drivers of HIV acquisition in African-American women related to mass incarceration: An agent-based modelling study. *BMC Public Health*. 2018;18(1):1-11. doi:10.1186/s12889-018-6304-x

13. Cauchemez S, Bhattarai A, Marchbanks TL, et al. Role of social networks in shaping disease transmission during a community outbreak of 2009 H1N1 pandemic influenza. *Proceedings of the National Academy of Sciences*. 2011;108(7):2825-2830. doi:10.1073/pnas.1008895108

14. Moran E, Kubale J, Noppert G, Malosh RE, Zelner JL. Inequality in acute respiratory infection outcomes in the United States: A review of the literature and its implications for public health policy

and practice. *medRxiv*. Published online April 26, 2020:2020.04.22.20069781. doi:10.1101/2020.04.22.20069781

15. Woolhouse M. How to make predictions about future infectious disease risks. *Philosophical Transactions of the Royal Society B: Biological Sciences*. 2011;366(1573):2045-2054. doi:10.1098/rstb.2010.0387

16. Zelner J, Riou J, Etzioni R, Gelman A. Accounting for uncertainty during a pandemic. *Patterns*. 2021;2(8):100310. doi:10.1016/j.patter.2021.100310

17. Ma KC, Menkir TF, Kissler SM, Grad YH, Lipsitch M. Modeling the impact of racial and ethnic disparities on COVID-19 epidemic dynamics. Schiffer JT, ed. *eLife*. 2021;10:e66601. doi:10.7554/eLife.66601

18. Coleman W. *Death Is a Social Disease: Public Health and Political Economy in Early Industrial France*. Univ of Wisconsin Pr; 1982.

19. Dubos RJ. Second Thoughts on the Germ Theory. *Scientific American*. 1955;192(5):31-35. Accessed September 4, 2020. http://www.jstor.org/stable/24944640

20. Dubos R, Dubos J. *The White Plague: Tuberculosis, Man, and Society*. Rutgers University Press; 1959.

21. Roberts SK. *Infectious Fear: Politics, Disease, and the Health Effects of Segregation*. University of North Carolina Press; 2009.

22. Johnson S. *The Ghost Map: The Story of London's Most Terrifying Epidemic--and How It Changed Science, Cities, and the Modern World*. Reprint edition. Riverhead Books; 2007.

23. Mojola SA. Fishing in dangerous waters: Ecology, gender and economy in HIV risk. *Social Science & Medicine*. 2011;72(2):149-156. doi:10.1016/j.socscimed.2010.11.006

24. Mojola SA, Wamoyi J. Contextual drivers of HIV risk among young African women. *J Int AIDS Soc*. 2019;22(Suppl Suppl 4). doi:10.1002/jia2.25302

25. Clouston SAP, Yukich J, Anglewicz P. Social inequalities in malaria knowledge, prevention and prevalence among children under 5 years old and women aged 15–49 in Madagascar. *Malar J*. 2015;14(1):499. doi:10.1186/s12936-015-1010-y

26. Faust C, Zelner J, Brasseur P, et al. Assessing drivers of full adoption of test and treat policy for malaria in Senegal. *American Journal of Tropical Medicine and Hygiene*. 2015;93(1):159-167. doi:10.4269/ajtmh.14-0595

27. Acevedo-Garcia D. Residential segregation and the epidemiology of infectious diseases. *Social Science & Medicine*. 2000;51(8):1143-1161. doi:10.1016/S0277-9536(00)00016-2

28. Clouston SAP, Rubin MS, Phelan JC, Link BG. A Social History of Disease: Contextualizing the Rise and Fall of Social Inequalities in Cause-Specific Mortality. *Demography*. 2016;53(5):1631-1656. doi:10.1007/s13524-016-0495-5

29. Bertozzi AL, Franco E, Mohler G, Short MB, Sledge D. The challenges of modeling and forecasting the spread of COVID-19. *PNAS*. 2020;117(29):16732-16738.